\documentclass[prl,amsmath,amssymb,superscriptaddress,twocolumn,floatfix]{revtex4-1}
\usepackage{graphicx}
\usepackage{dcolumn}
\usepackage{bm}
\usepackage{tabularx}

\def\bscco{Bi$_2$Sr$_2$CaCu$_2$O$_{8+y}$}

\def\lsco{La$_{2-x}$Sr$_x$CuO$_4$}
\def\lbco{La$_{2-x}$Ba$_x$CuO$_4$}

\def\ybco{YBa$_2$Cu$_3$O$_{6+y}$}
\def\qstm{${\bf q}_{\rm STM}$}
\def\qco{${\bf q}_{\rm co}$}
\def\qso{${\bf q}_{\rm so}$}

\begin{document}


\title{Spin-stripe density varies linearly with hole content in single-layer Bi2201 cuprate}


\author{M. Enoki}
\affiliation{Department of Physics, Tohoku University, Sendai 980-8578, Japan}
\email[]{enoki@post.matsc.kyutech.ac.jp}
\author{M. Fujita}
\affiliation{Institute for Materials Research, Tohoku University, Sendai 980-8577, Japan}
\author{T. Nishizaki}
\affiliation{Institute for Materials Research, Tohoku University, Sendai 980-8577, Japan}
\author{S. Iikubo}
\affiliation{Kyushu Institute of Technology, Kitakyushu 808-0196, Japan}
\author{D. K. Singh}
\affiliation{NIST Center for Neutron Research, National Institute of Standards and Technology, Gaithersburg, Maryland 20899, USA}
\affiliation{Department of Materials Science and Engineering, University of Maryland, College Park, Maryland 20899, USA}
\author{S. Chang}
\affiliation{NIST Center for Neutron Research, National Institute of Standards and Technology, Gaithersburg, Maryland 20899, USA}
\author{J. M. Tranquada}
\affiliation{Condensed Matter Physics \&\ Materials Science Dept., Brookhaven National Laboratory, Upton, New York 11973-5000, USA}
\author{K. Yamada}
\affiliation{WPI Advanced Institute for Materials Research, Tohoku University, Sendai 980-8577, Japan}


\date{\today}

\begin{abstract}
We have performed inelastic neutron scattering measurements on the single-layer cuprate Bi$_{2+x}$Sr$_{2-x}$CuO$_{6+y}$ (Bi2201) with $x$=0.2, 0.3, 0.4 and 0.5, a doping range that spans the spin-glass (SG) to superconducting (SC) phase boundary. The doping evolution of low energy spin fluctuations ($\lesssim11$~meV) was found to be characterized by a change of incommensurate modulation wave vector from the tetragonal [110] to [100]/[010] directions, while maintaining a linear relation between the incommensurability and the hole concentration, $\delta\approx p$.  In the SC regime, the spectral weight is strongly suppressed below $\sim4$~meV.  Similarities and differences in the spin correlations between Bi2201 and the prototypical single-layer system La$_{2-x}$Sr$_x$CuO$_4$ are discussed. 

\end{abstract}


\maketitle


The relevance of charge and spin stripes to the phenomenology of hole-doped cuprate superconductors has been gaining currency in recent years.  For example, intriguing similarities in the transport properties of stripe-ordered cuprates and \ybco\ (YBCO) in a high magnetic field have been demonstrated \cite{chan10,lebo11}.  In fact, direct evidence for field-induced charge-stripe order in YBCO was recently obtained in a nuclear magnetic resonance study \cite{wu11}.  The impact of stripe order on the Fermi surface \cite{mill07,yao11} has been proposed as one possible explanation for the appearance of quantum oscillations \cite{lebo07,seba08} (though there are challenges with the simplest approaches \cite{vojt12}).   There is also evidence of closely related nematic order \cite{kive98} from Nernst-effect measurements on YBCO \cite{daou10} as well as from spectroscopic imaging of \bscco\ with scanning tunneling microscopy (STM) \cite{lawl10,mesa11}.

Some of the most stimulating evidence comes from real-space imaging of electronic modulations by STM in \bscco\ \cite{howa03b,kohs07,park10} and in Bi$_{2-x}$Pb$_x$Sr$_{2-z}$La$_z$CuO$_{6+y}$ \cite{wise08}.  These short-range correlations are found to have a period of approximately $4a$, where $a=3.8$~\AA\ is the lattice spacing along a Cu-O bond direction.  The corresponding wave vector of the modulations, \qstm, is $(\frac14,0,0)$ in reciprocal lattice units ($2\pi/a$); $q_{\rm STM}$ is observed to decrease with doping, varying in the range of 0.3 to 0.15 \cite{park10,wise08}.  

Identifying \qstm\ with the wave vector \qco\ associated with charge stripe order in cuprates such as \lbco\ \cite{fuji04} leads to a conundrum, as $q_{\rm co}$ {\it grows} with doping (at least for hole concentrations $p \lesssim 1/8$ \cite{huck11}), opposite to the behavior of \qstm.  When spin stripe order also occurs, antiferromagnetic spin correlations are modulated at ${\bf q}_{\rm so} = \frac12{\bf q}_{\rm co}$.  It is often possible to observe incommensurate (IC) spin fluctuations split about the antiferromagnetic wave vector ${\bf Q}_{\rm AF}$ by ${\bf q}_\delta\approx{\bf q}_{\rm so}$ even when there is no significant stripe order, as in \lsco\ (LSCO) \cite{yama98a,birg06}.   Among possible resolutions of the conundrum, it might be that the nature of stripe correlations is not universal among different cuprate families, or that \qstm\ measures something complementary to \qco.

In this Letter, we present the results of inelastic neutron scattering measurements of low-energy spin excitations in the system Bi$_{2+x}$Sr$_{2-x}$CuO$_{6+y}$ (Bi2201), demonstrating that $\delta=|{\bf q}_\delta|\approx p$ for $0.01\lesssim p\lesssim0.12$.  This behavior is remarkably similar to that of \lsco, even including the rotation of \qso\ by $45^\circ$ for $p\lesssim0.06$ \cite{yama98a,birg06}.  These results provide strong circumstantial evidence that \qstm\ does not correspond to \qco; instead, it more likely corresponds to a nesting of antinodal states close to $2k_{\rm F}$, where $k_{\rm F}$ is the nominal Fermi wave vector \cite{shen05, wise08}.    This is not incompatible with a stripe origin, but would involve modulations along the charge stripes rather than perpendicular to them.

While several variants of Bi2201 have been studied in the literature, we chose to work with Bi$_{2+x}$Sr$_{2-x}$CuO$_{6+y}$ because it is possible both to vary the hole concentration in a controlled fashion and to grow sufficiently large crystals with the floating-zone technique, as previously demonstrated by Luo {\it et al.}\ \cite{luo07}.
We prepared single crystals of Bi2201 with $x$=0.2, 0.3, 0.4, and 0.5.
The actual concentrations of Bi and Sr were determined by inductively-coupled plasma (ICP) atomic emission spectroscopy (AES), and the hole densities were determined by measurements of the Hall coefficient at 200~K, following \cite{ando00};
the results are listed in Table 1. The correspondence between $p$ and $x$ is consistent with the previously reported results based on measurements of the Fermi-surface volume by angle-resolved photoemission spectroscopy \cite{pan09}.  Based on magnetic susceptibility measurements, spin-glass-like behavior was observed below 3~K for $x=0.4$ \cite{enok11} and below 4~K for $x=0.5$; neither magnetic order nor diamagnetism were detected above 2~K in the $x=0.3$ and 0.2 samples.  According to Luo {\it et al.}\ \cite{luo07}, the superconducting transition temperature, $T_c$, is $\sim1$~K at $x=0.2$, rising up to a maximum of 9~K at $x=0.05$, as shown in Fig.~\ref{delta}(b).  The reduced $T_c$ in this system compared to La substitution for Sr is likely associated with structural disorder \cite{hobo09}.

  \begin{table}[t]
    \begin{center}
\newcolumntype{C}{>{\centering\arraybackslash}X}
\caption{Characterizations of the Bi$_{2+x}$Sr$_{2-x}$CuO$_{6+y}$ crystals. Elemental concentrations were determined by ICP-AES and hole concentration $p$ was determined from Hall effect measurements.
 }
    \begin{tabularx}{82mm}{|C||C|C|C|C|}
    \hline
      $x$ & Bi & Sr & Cu & $p$\\ \hline\hline
      0.2 & 2.173(1) & 1.825(1) & 0.989(2) & 0.12(1)\\ \hline
      0.3 & 2.282(3) & 1.717(2) & 0.992(4) & 0.09(1)\\ \hline
      0.4 & 2.376(1) & 1.619(1) & 0.992(2) & 0.06(1)\\ \hline
      0.5 & - & - & - & 0.01(1)\\ \hline
    \end{tabularx}
    \end{center}
  \end{table}


Most of the inelastic neutron scattering measurements were performed on thermal triple-axis spectrometer TOPAN 
installed at reactor JRR-3, Japan Atomic Energy Agency (JAEA).
The typical collimator selections were $50'$-$100'$-Sample-$60'$-$180'$, and the final energy was fixed at 14.7~meV. To reduce contamination from high-energy neutrons, a sapphire crystal was placed before the sample. A pyrolytic graphite filter was placed after the sample to eliminate higher-order neutrons. 
Additional measurements below 4~meV were performed at the cold neutron triple-axis spectrometers HER installed in the Guide Hall of JRR-3 and SPINS at the NIST Center for Neutron Research. 
For each composition, a couple of single crystals with total mass of 10--15 grams were coaligned and positioned so that the scattering plane corresponds to $(h,k,0)$.  Some results for the $x=0.4$ crystal were reported previously \cite{enok11}. 

For consistency, we will continue to index the scattering in terms of a tetragonal unit cell with $a_t=b_t\approx3.81$~\AA, although the symmetry is actually orthorhombic, with in-plane basis vectors along $[1\bar{1}0]$ and $[110]$ corresponding to $a_o$ and $b_o$, respectively.  Although we cannot resolve the very small orthorhombic strain, we can distinguish the $b_o^*$ direction by the presence of superlattice peaks (at $\sim 0.2 b_o^*$) corresponding to the modulation of the BiO layers.   We find that $b_o^*$ runs in a unique direction in each crystal ({\it i.e.}, there is little, if any, twinning), and we will see that this results in a unique orientation of the IC spin fluctuations in the more underdoped crystals.


\begin{figure}[t]
\begin{center}
\includegraphics[bb=0 0 837 792, width=3.3in]{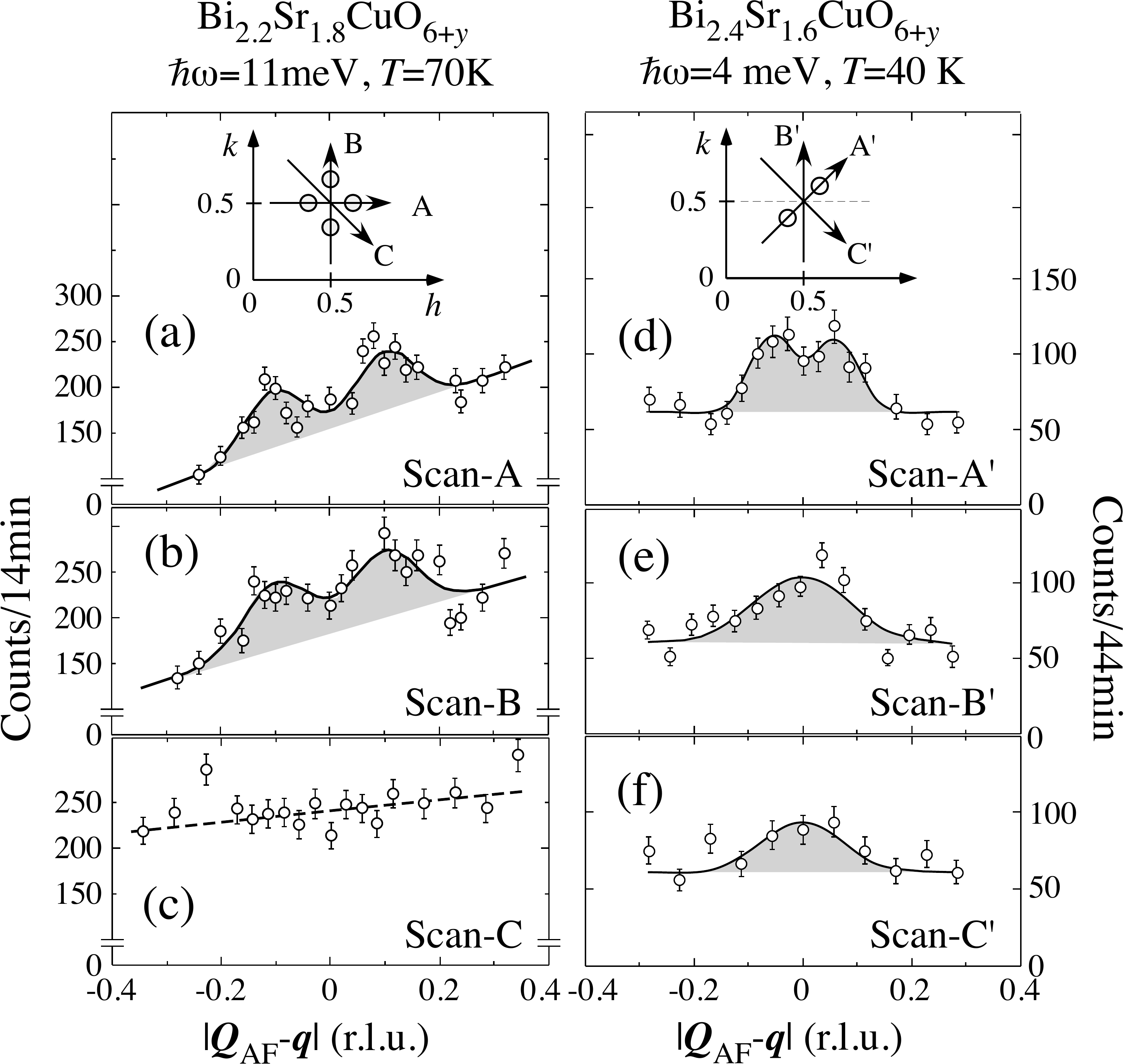}
\end{center}
\caption{Inelastic neutron scattering spectra in Bi$_{2.2}$Sr$_{1.8}$CuO$_{6+y}$ $x$=0.2 at 11 meV, 70 K (a)-(c) and $x$=0.4 at 4 meV, 40 K (d)-(f). Insets show the IC peaks geometry and the scan trajectory. IC peaks are shown in (a)[100] and (b)[010] direction in $x$=0.2, and (d)[110] direction in $x$=0.4 sample, respectively.}
\label{INS}
\end{figure}

Inelastic neutron-scattering spectra for the $x=0.2$ sample ($p=0.12$) obtained for an excitation energy of $\hbar\omega= 11$~meV and a temperature of $T= 70$~K are shown in Fig.~\ref{INS}(a)-(c).  Scans A and B exhibit IC peaks split about ${\bf Q}_{\rm AF}$ in the [100] and [010] directions, while the transverse scan C shows no structure.  The pattern is identical to that observed in the superconducting phase of LSCO \cite{yama98} and twinned YBCO \cite{dai01,haug10}; however, the intensity at this and lower energies is weak compared to that from LSCO for the same $p$ and mass, measured under identical experimental setups.

Related scans for the $x=0.4$ sample ($p=0.06$) are shown in Fig.~\ref{INS}(d)-(f); these were measured at $\hbar\omega=4$~meV and $T=40$~K using the SPINS spectrometer with $E_f=5$~meV.  Here we see that IC peaks are only in scan A', which is along $b_o^*$, with no IC peaks along scan C', in the direction of $a_o^*$.  Similar scans at $\hbar\omega=1$~meV are reported in \cite{enok11}, where the intensity is shown to fall off with temperature in a fashion consistent with magnetic correlations.  An earlier study demonstrated that the signal falls off in higher Brillouin zones, as expected for a magnetic form factor \cite{enok10}.  The observation of a longitudinal IC splitting along a unique orthorhombic axis corresponds perfectly with the behavior previously found in the spin-glass phase of LSCO \cite{waki00,birg06}.  From the unique orientation we infer that static order is likely, and it is strongly indicated by bulk susceptibility measurements to occur below 3~K \cite{enok11}; however, we were not able to detect IC peaks in elastic scattering for any of the samples.  Of course, there is a substantial background in the elastic channel from nuclear diffuse scattering resulting from structural disorder, and that limits the sensitivity.

\begin{figure}[t]
\begin{center}
\includegraphics[bb=0 0 485 712, width=6.7cm]{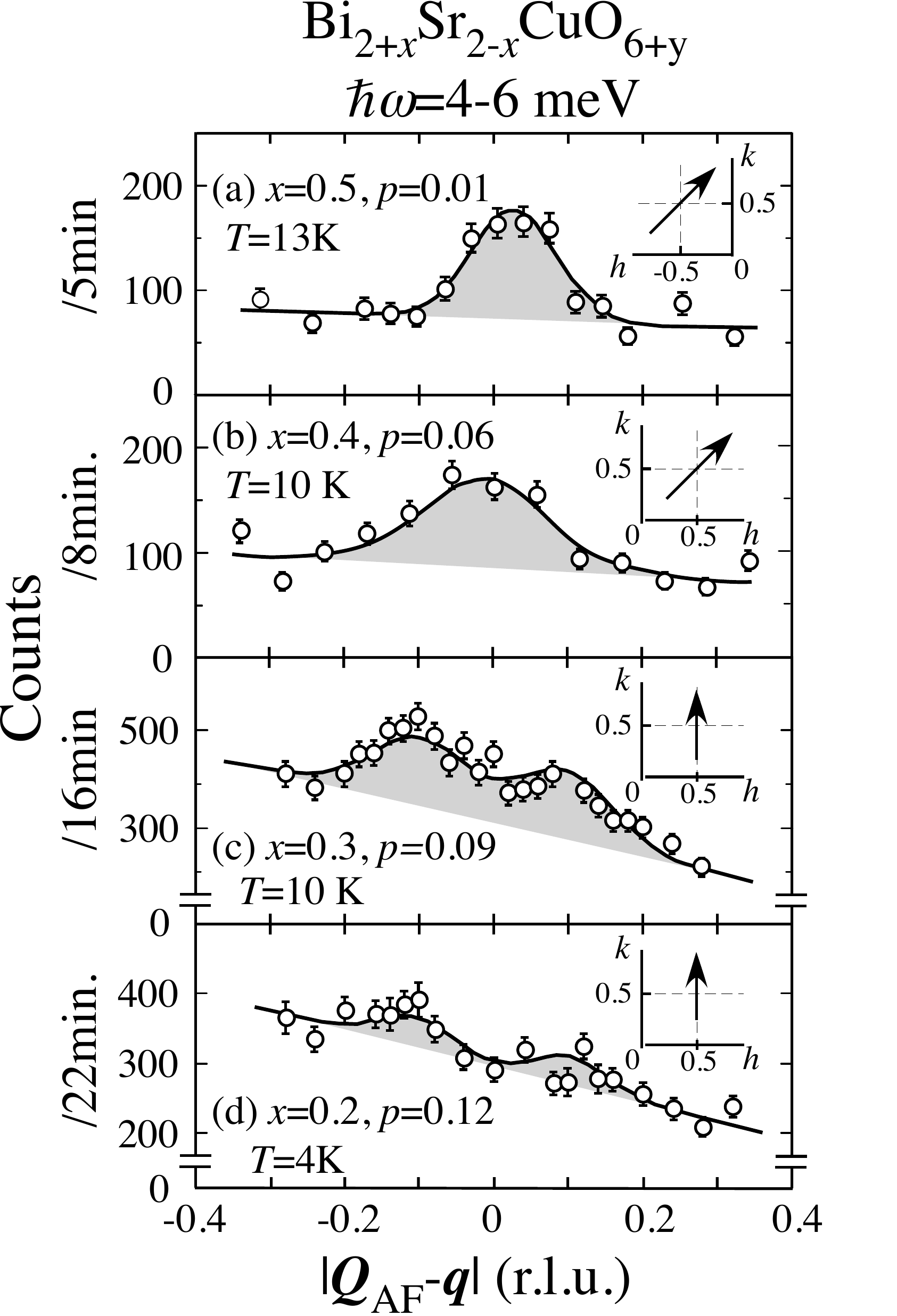}
\end{center}
\caption{Inelastic neutron scattering spectra at 4-6 meV and low temperatures in Bi$_{2.2}$Sr$_{1.8}$CuO$_{6+y}$ with $x$=(a) 0.5, (b) 0.4, (c) 0.3 and (d) 0.2. Spectra were measured along [110] and [010] directions for (a,b) and (c,d), respectively.}
\label{doping}
\end{figure}

To illustrate the variation of the spin correlations for all four of our samples, Fig.~\ref{doping} shows representative scans through ${\bf Q}_{\rm AF}$ obtained with thermal neutrons at excitation energies in the range of 4--6 meV.  For $x=0.3$, the orientation of the IC peaks is the same as for $x=0.2$, but the splitting $\delta$ is slightly smaller.  For both $x=0.4$ and 0.5, the instrumental resolution is too broad to resolve IC peaks, but the narrower width for $x=0.5$ suggests a smaller splitting.

For quantitative analysis, we model the scattered intensity $I({\bf Q},\omega)\sim\chi''({\bf Q},\omega)(1-e^{-\hbar\omega/{\it k}_{\rm B}{\it T}})^{-1}$ with the formula,
%
\begin{eqnarray*}
\chi''({\bf Q},\omega)=\chi''(\omega)
\exp\left[-\ln(2)
({\bf Q}-{\bf Q}_{\rm AF}\pm{\bf q}_{\delta})^2/\kappa^2 \right].
\end{eqnarray*}
%
$\chi^{\prime\prime}(\omega)$ and $\kappa$ correspond to the local spin susceptibility and the peak-width (half-width at half-maximum), respectively.   In fitting the data, the model intensity was convolved with the instrumental resolution function and a linear background was included.
For the $x=0.2$ and 0.3 samples, we defined ${\bf q}_\delta$ to include $(\delta, 0, 0)$ and $(0, \delta, 0)$.  For $x=0.4$ and 0.5, we set ${\bf q}_\delta=(\delta/\sqrt{2}, \delta/\sqrt{2}, 0)$; fitting the high-resolution data of Fig.~\ref{INS}(d) yielded $\delta=0.057(5)$. 
Since the fitted values of $\kappa$ for $x=0.4$, 0.3, and 0.2 in Fig 1(d) and Fig 2(c,d) were comparable ($\sim0.04$ r.l.u.) , we fitted the $x=0.5$ data with $\kappa$ fixed at 0.04 and with the assumption of IC peaks oriented as for $x=0.4$. 

\begin{figure}[t]
\begin{center}
\includegraphics[bb=0 0 365 485, width=6.8cm]{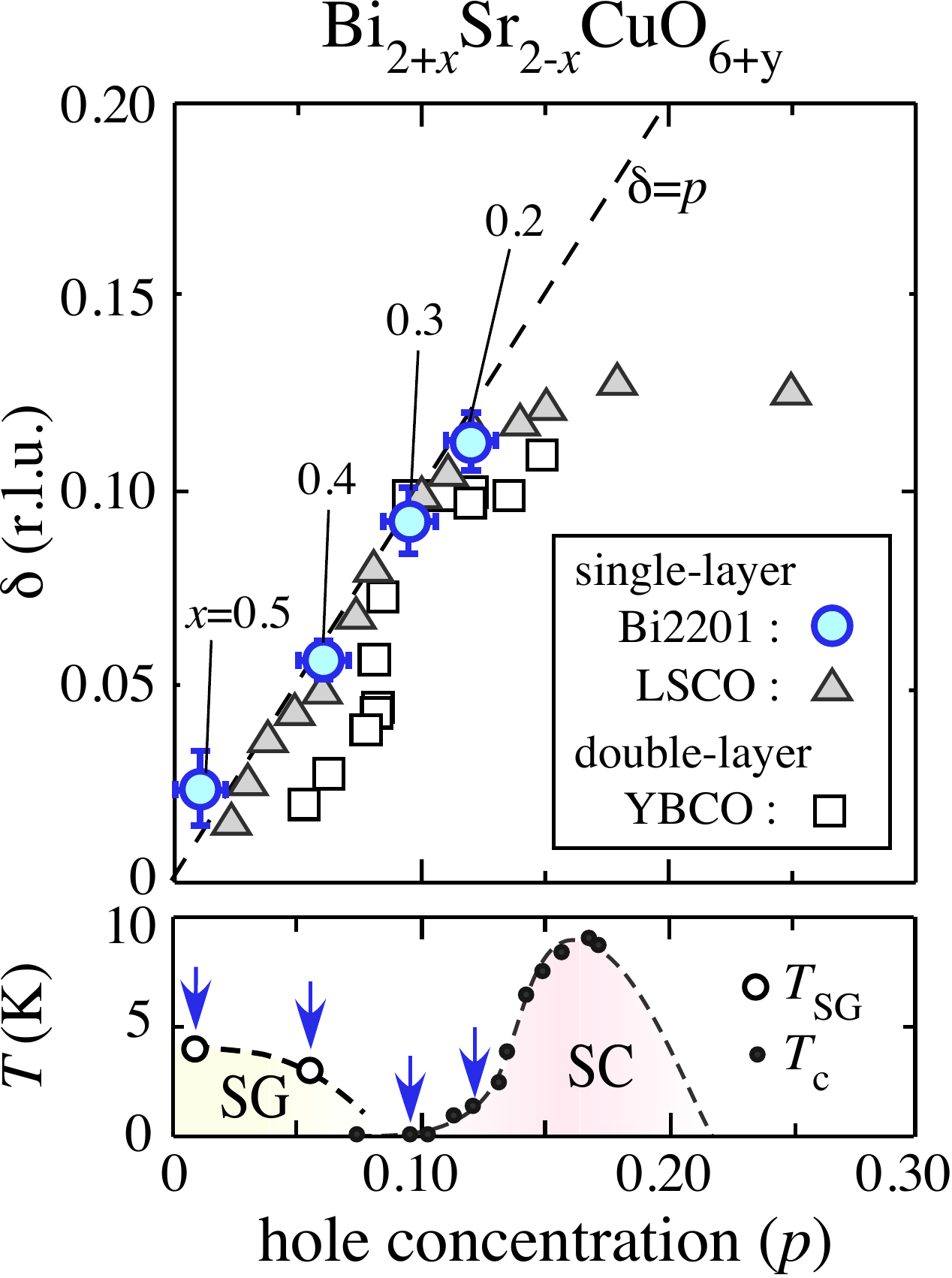}
\end{center}
\caption{(Color online) (upper panel) Hole concentration dependence of the incommensurability $\delta$ of low-energy spin fluctuations in Bi$_{2+x}$Sr$_{2-x}$CuO$_{6+y}$ (blue circles) compared with results for LSCO (gray triangles) \cite{yama98,waki00,fuji02c} and YBCO (open squares) \cite{dai01,haug10} 
Dashed line represents $\delta$=$p$.
(lower panel) Electronic phase diagram of Bi$_{2+x}$Sr$_{2-x}$CuO$_{6+y}$. Spin-glass transition temperature is plotted by open circles. $T_c$ data (filled circles) are from \cite{luo08}. Dashed lines are guide to the eye. }
\label{delta}
\end{figure}

The values of $\delta$ obtained for Bi2201 from the fitting are plotted as a function of $p$ in the top panel of Fig.~\ref{delta}, where they are compared with results for  LSCO \cite{yama98,waki00,fuji02c} and YBCO \cite{dai01,haug10}.  We find that $\delta\approx p$ in Bi2201, which appears to be quantitatively identical to LSCO and qualitatively similar to YBCO.  Comparing with the spin-glass and nominal superconducting transitions indicated at the bottom of Fig.~\ref{delta}, the rotation of ${\bf q}_\delta$ occurs between the spin-glass and superconducting phases, just as in LSCO \cite{birg06}.

The degree of similarity between Bi2201 and LSCO is a bit surprising, given that photoemission studies have indicated significant differences \cite{hash08}.  In particular, the chemical potential in LSCO remains rather constant with doping for $0\le p\lesssim0.12$, while it shifts downward linearly with doping in Bi2201 \cite{hash08}.  There have been variety of models proposed to explain the doping dependence of $\delta$ in cuprates \cite{sush09,seib11}, not all of which involve stripes; nevertheless, if stripes are involved, then it is reasonable to expect that ${\bf q}_{\rm co} \approx 2{\bf q}_\delta$, and hence $|{\bf q}_{\rm co}|\approx 2p$ for Bi2201 with $p\lesssim0.12$ based on the present results.

\begin{figure}[t]
\begin{center}
\includegraphics[bb=0 0 429 669, width=6.8cm]{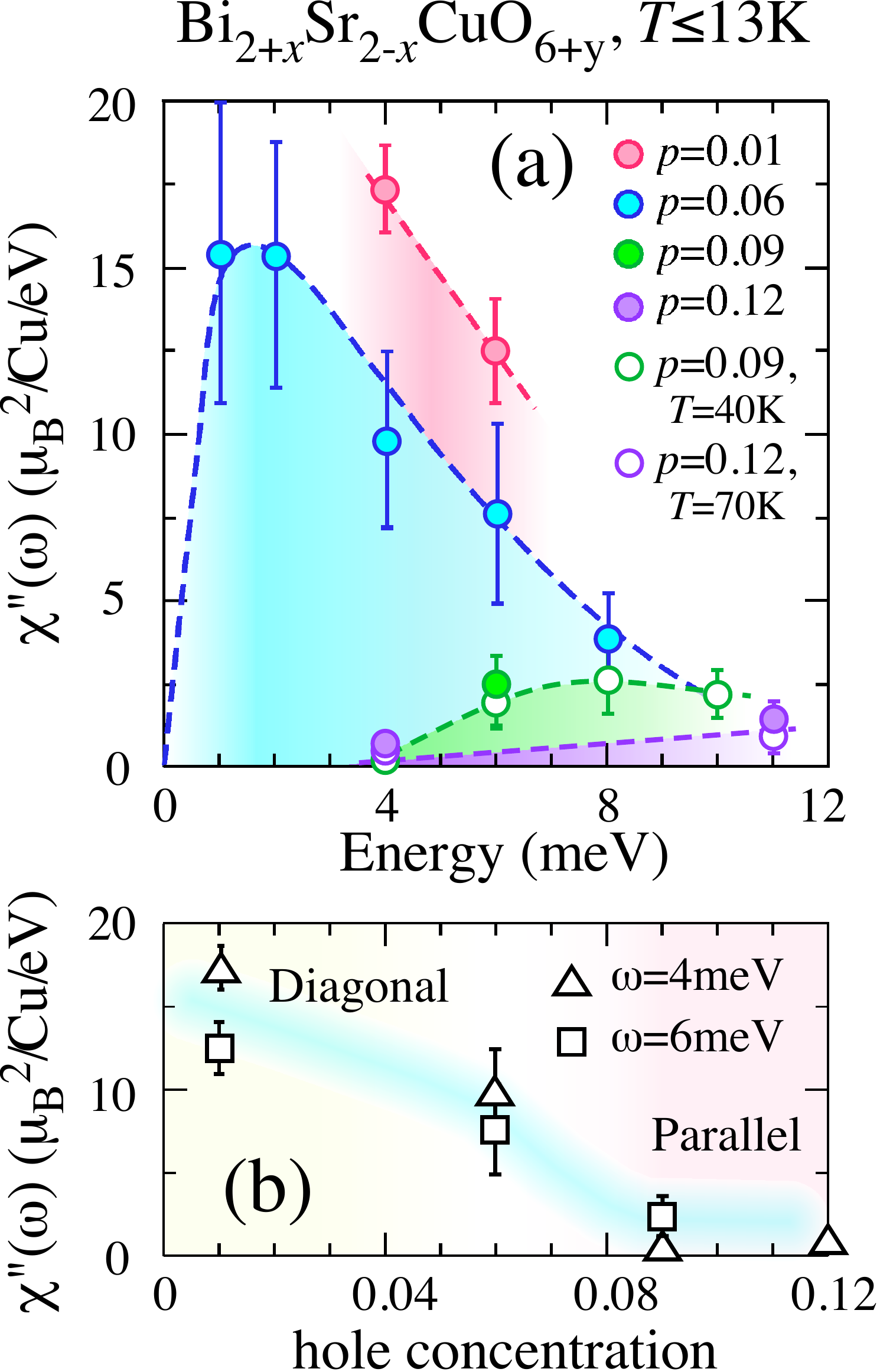}
\end{center}
\caption{(Color online) (a) Energy dependence of the local spin susceptibility $\chi^{\prime\prime}(\omega)$ for Bi$_{2+x}$Sr$_{2-x}$CuO$_{6+y}$ with $x$=0.2, 0.3, 0.4 and 0.5. Results at low temperatures below 13 K and high temperatures of 70 K for $x$=0.2 and 40K for $x$=0.3 are plotted by closed and open circles, respectively. (b) Hole concentration dependence of $\chi^{\prime\prime}(\omega)$ at 4 meV (triangles) and 6 meV (squares). Dashed lines in (a) and broad line in (b) are guides to the eye. }
\label{tdep}
\end{figure}


The experimental frequency dependence of the local susceptibility $\chi''(\omega)$ is plotted in Fig.~\ref{tdep}(a).  For the two samples near the spin-glass regime ($p=0.01$ and 0.06), the low-temperature $\chi''(\omega)$ is large at low energy, consistent with proximity to an ordered state.  For the more highly-doped samples ($p=0.09$ and 0.12), $\chi''(\omega)$ is strongly reduced at low energy, and there is not much change when the temperature is raised somewhat (open symbols).   The doping trend for the low-energy weight is illustrated in Fig.~\ref{tdep}(b).  While it is qualitatively consistent with behavior observed in cuprates, the rapid loss of low-energy weight in this doping range is different from LSCO \cite{lee00}.  In particular, the results indicate that the system is far from static spin order of the type associate with the 1/8 anomaly in LSCO. 

Our result $\delta\approx p$ in Bi2201 provides further evidence of universal behavior of spin correlations in the cuprates.  The implication for possible coexisting charge modulations has a clear implication for the interpretation of STM studies.  The doping dependence of ${\bf q}_{\rm STM}$ is not consistent with the simplest stripe interpretation.  The electronic modulations are more likely due to $2k_{\rm F}$-like modulations associated with the large antinodal density of states \cite{wise08,shen05}.  We are aware that ${\bf q}_{\rm STM}$ does not precisely match $2k_{\rm F}$ measured by photoemssion \cite{gweo11}; however, this is not a problem as the STM modulations have maximum amplitude at bias voltages comparable to the pseudogap energy, so that the ${\bf q}_{\rm STM}$ need not be determined by states precisely at the Fermi level.   It may be relevant that the antinodal pseudogap energy is also the scale on which antiferromagnetic spin fluctuations become strongly damped \cite{stoc10}.  Hence, the low-energy spin fluctuations and the STM modulations appear to detect different electronic features with distinct energy scales. 

We would like to thank N. Kobayashi, T. Adachi, K. Kudo, R. He, Y. Koike, M. Mori, M. Hiraka, O. Sushkov and H. Yamase for the fruitful discussions. ME is supported by Technology and Global COE Program "Materials Integration, Tohoku University", MEXT Japan. MF is supported by Grant-in-Aid for Scientific Research (B) (23340093). JMT is supported at Brookhaven by the Office of Basic Energy Sciences, Division of Materials Science and Engineering, U.S. Department of Energy (DOE), under Contract No. DE-AC02-98CH10886.  SPINS at NCNR is supported by the National Science Foundation under Agreement No.\ DMR-0454672. The work at JRR-3 was partially performed under the Common-Use Facility Program of JAEA and joint research program of ISSP, the University of Tokyo.

\bibliography{lno,theory}

\begin{thebibliography}{42}%
\makeatletter
\providecommand \@ifxundefined [1]{%
 \@ifx{#1\undefined}
}%
\providecommand \@ifnum [1]{%
 \ifnum #1\expandafter \@firstoftwo
 \else \expandafter \@secondoftwo
 \fi
}%
\providecommand \@ifx [1]{%
 \ifx #1\expandafter \@firstoftwo
 \else \expandafter \@secondoftwo
 \fi
}%
\providecommand \natexlab [1]{#1}%
\providecommand \enquote  [1]{``#1''}%
\providecommand \bibnamefont  [1]{#1}%
\providecommand \bibfnamefont [1]{#1}%
\providecommand \citenamefont [1]{#1}%
\providecommand \href@noop [0]{\@secondoftwo}%
\providecommand \href [0]{\begingroup \@sanitize@url \@href}%
\providecommand \@href[1]{\@@startlink{#1}\@@href}%
\providecommand \@@href[1]{\endgroup#1\@@endlink}%
\providecommand \@sanitize@url [0]{\catcode `\\12\catcode `\$12\catcode
  `\&12\catcode `\#12\catcode `\^12\catcode `\_12\catcode `\%12\relax}%
\providecommand \@@startlink[1]{}%
\providecommand \@@endlink[0]{}%
\providecommand \url  [0]{\begingroup\@sanitize@url \@url }%
\providecommand \@url [1]{\endgroup\@href {#1}{\urlprefix }}%
\providecommand \urlprefix  [0]{URL }%
\providecommand \Eprint [0]{\href }%
\providecommand \doibase [0]{http://dx.doi.org/}%
\providecommand \selectlanguage [0]{\@gobble}%
\providecommand \bibinfo  [0]{\@secondoftwo}%
\providecommand \bibfield  [0]{\@secondoftwo}%
\providecommand \translation [1]{[#1]}%
\providecommand \BibitemOpen [0]{}%
\providecommand \bibitemStop [0]{}%
\providecommand \bibitemNoStop [0]{.\EOS\space}%
\providecommand \EOS [0]{\spacefactor3000\relax}%
\providecommand \BibitemShut  [1]{\csname bibitem#1\endcsname}%
\let\auto@bib@innerbib\@empty
\bibitem [{\citenamefont {Chang}\ \emph {et~al.}(2010)\citenamefont {Chang},
  \citenamefont {Daou}, \citenamefont {Proust}, \citenamefont {LeBoeuf},
  \citenamefont {Doiron-Leyraud}, \citenamefont {Lalibert\'e}, \citenamefont
  {Pingault}, \citenamefont {Ramshaw}, \citenamefont {Liang}, \citenamefont
  {Bonn}, \citenamefont {Hardy}, \citenamefont {Takagi}, \citenamefont
  {Antunes}, \citenamefont {Sheikin}, \citenamefont {Behnia},\ and\
  \citenamefont {Taillefer}}]{chan10}%
  \BibitemOpen
  \bibfield  {author} {\bibinfo {author} {\bibfnamefont {J.}~\bibnamefont
  {Chang}}, \bibinfo {author} {\bibfnamefont {R.}~\bibnamefont {Daou}},
  \bibinfo {author} {\bibfnamefont {C.}~\bibnamefont {Proust}}, \bibinfo
  {author} {\bibfnamefont {D.}~\bibnamefont {LeBoeuf}}, \bibinfo {author}
  {\bibfnamefont {N.}~\bibnamefont {Doiron-Leyraud}}, \bibinfo {author}
  {\bibfnamefont {F.}~\bibnamefont {Lalibert\'e}}, \bibinfo {author}
  {\bibfnamefont {B.}~\bibnamefont {Pingault}}, \bibinfo {author}
  {\bibfnamefont {B.~J.}\ \bibnamefont {Ramshaw}}, \bibinfo {author}
  {\bibfnamefont {R.}~\bibnamefont {Liang}}, \bibinfo {author} {\bibfnamefont
  {D.~A.}\ \bibnamefont {Bonn}}, \bibinfo {author} {\bibfnamefont {W.~N.}\
  \bibnamefont {Hardy}}, \bibinfo {author} {\bibfnamefont {H.}~\bibnamefont
  {Takagi}}, \bibinfo {author} {\bibfnamefont {A.~B.}\ \bibnamefont {Antunes}},
  \bibinfo {author} {\bibfnamefont {I.}~\bibnamefont {Sheikin}}, \bibinfo
  {author} {\bibfnamefont {K.}~\bibnamefont {Behnia}}, \ and\ \bibinfo {author}
  {\bibfnamefont {L.}~\bibnamefont {Taillefer}},\ }\href@noop {} {\bibfield
  {journal} {\bibinfo  {journal} {Phys. Rev. Lett.}\ }\textbf {\bibinfo
  {volume} {104}},\ \bibinfo {pages} {057005} (\bibinfo {year}
  {2010})}\BibitemShut {NoStop}%
\bibitem [{\citenamefont {LeBoeuf}\ \emph {et~al.}(2011)\citenamefont
  {LeBoeuf}, \citenamefont {Doiron-Leyraud}, \citenamefont {Vignolle},
  \citenamefont {Sutherland}, \citenamefont {Ramshaw}, \citenamefont
  {Levallois}, \citenamefont {Daou}, \citenamefont {Lalibert\'e}, \citenamefont
  {Cyr-Choini\`ere}, \citenamefont {Chang}, \citenamefont {Jo}, \citenamefont
  {Balicas}, \citenamefont {Liang}, \citenamefont {Bonn}, \citenamefont
  {Hardy}, \citenamefont {Proust},\ and\ \citenamefont {Taillefer}}]{lebo11}%
  \BibitemOpen
  \bibfield  {author} {\bibinfo {author} {\bibfnamefont {D.}~\bibnamefont
  {LeBoeuf}}, \bibinfo {author} {\bibfnamefont {N.}~\bibnamefont
  {Doiron-Leyraud}}, \bibinfo {author} {\bibfnamefont {B.}~\bibnamefont
  {Vignolle}}, \bibinfo {author} {\bibfnamefont {M.}~\bibnamefont
  {Sutherland}}, \bibinfo {author} {\bibfnamefont {B.~J.}\ \bibnamefont
  {Ramshaw}}, \bibinfo {author} {\bibfnamefont {J.}~\bibnamefont {Levallois}},
  \bibinfo {author} {\bibfnamefont {R.}~\bibnamefont {Daou}}, \bibinfo {author}
  {\bibfnamefont {F.}~\bibnamefont {Lalibert\'e}}, \bibinfo {author}
  {\bibfnamefont {O.}~\bibnamefont {Cyr-Choini\`ere}}, \bibinfo {author}
  {\bibfnamefont {J.}~\bibnamefont {Chang}}, \bibinfo {author} {\bibfnamefont
  {Y.~J.}\ \bibnamefont {Jo}}, \bibinfo {author} {\bibfnamefont
  {L.}~\bibnamefont {Balicas}}, \bibinfo {author} {\bibfnamefont
  {R.}~\bibnamefont {Liang}}, \bibinfo {author} {\bibfnamefont {D.~A.}\
  \bibnamefont {Bonn}}, \bibinfo {author} {\bibfnamefont {W.~N.}\ \bibnamefont
  {Hardy}}, \bibinfo {author} {\bibfnamefont {C.}~\bibnamefont {Proust}}, \
  and\ \bibinfo {author} {\bibfnamefont {L.}~\bibnamefont {Taillefer}},\
  }\href@noop {} {\bibfield  {journal} {\bibinfo  {journal} {Phys. Rev. B}\
  }\textbf {\bibinfo {volume} {83}},\ \bibinfo {pages} {054506} (\bibinfo
  {year} {2011})}\BibitemShut {NoStop}%
\bibitem [{\citenamefont {Wu}\ \emph {et~al.}(2011)\citenamefont {Wu},
  \citenamefont {Mayaffre}, \citenamefont {Kramer}, \citenamefont {Horvatic},
  \citenamefont {Berthier}, \citenamefont {Hardy}, \citenamefont {Liang},
  \citenamefont {Bonn},\ and\ \citenamefont {Julien}}]{wu11}%
  \BibitemOpen
  \bibfield  {author} {\bibinfo {author} {\bibfnamefont {T.}~\bibnamefont
  {Wu}}, \bibinfo {author} {\bibfnamefont {H.}~\bibnamefont {Mayaffre}},
  \bibinfo {author} {\bibfnamefont {S.}~\bibnamefont {Kramer}}, \bibinfo
  {author} {\bibfnamefont {M.}~\bibnamefont {Horvatic}}, \bibinfo {author}
  {\bibfnamefont {C.}~\bibnamefont {Berthier}}, \bibinfo {author}
  {\bibfnamefont {W.~N.}\ \bibnamefont {Hardy}}, \bibinfo {author}
  {\bibfnamefont {R.}~\bibnamefont {Liang}}, \bibinfo {author} {\bibfnamefont
  {D.~A.}\ \bibnamefont {Bonn}}, \ and\ \bibinfo {author} {\bibfnamefont
  {M.-H.}\ \bibnamefont {Julien}},\ }\href@noop {} {\bibfield  {journal}
  {\bibinfo  {journal} {Nature}\ }\textbf {\bibinfo {volume} {477}},\ \bibinfo
  {pages} {191} (\bibinfo {year} {2011})}\BibitemShut {NoStop}%
\bibitem [{\citenamefont {Millis}\ and\ \citenamefont {Norman}(2007)}]{mill07}%
  \BibitemOpen
  \bibfield  {author} {\bibinfo {author} {\bibfnamefont {A.~J.}\ \bibnamefont
  {Millis}}\ and\ \bibinfo {author} {\bibfnamefont {M.~R.}\ \bibnamefont
  {Norman}},\ }\href@noop {} {\bibfield  {journal} {\bibinfo  {journal} {Phys.
  Rev. B}\ }\textbf {\bibinfo {volume} {76}},\ \bibinfo {eid} {220503(R)}
  (\bibinfo {year} {2007})}\BibitemShut {NoStop}%
\bibitem [{\citenamefont {Yao}\ \emph {et~al.}(2011)\citenamefont {Yao},
  \citenamefont {Lee},\ and\ \citenamefont {Kivelson}}]{yao11}%
  \BibitemOpen
  \bibfield  {author} {\bibinfo {author} {\bibfnamefont {H.}~\bibnamefont
  {Yao}}, \bibinfo {author} {\bibfnamefont {D.-H.}\ \bibnamefont {Lee}}, \ and\
  \bibinfo {author} {\bibfnamefont {S.}~\bibnamefont {Kivelson}},\ }\href@noop
  {} {\bibfield  {journal} {\bibinfo  {journal} {Phys. Rev. B}\ }\textbf
  {\bibinfo {volume} {84}},\ \bibinfo {pages} {012507} (\bibinfo {year}
  {2011})}\BibitemShut {NoStop}%
\bibitem [{\citenamefont {LeBoeuf}\ \emph {et~al.}(2007)\citenamefont
  {LeBoeuf}, \citenamefont {Doiron-Leyraud}, \citenamefont {Levallois},
  \citenamefont {Daou}, \citenamefont {Bonnemaison}, \citenamefont {Hussey},
  \citenamefont {Balicas}, \citenamefont {Ramshaw}, \citenamefont {Liang},
  \citenamefont {Bonn}, \citenamefont {Hardy}, \citenamefont {Adachi},
  \citenamefont {Proust},\ and\ \citenamefont {Taillefer}}]{lebo07}%
  \BibitemOpen
  \bibfield  {author} {\bibinfo {author} {\bibfnamefont {D.}~\bibnamefont
  {LeBoeuf}}, \bibinfo {author} {\bibfnamefont {N.}~\bibnamefont
  {Doiron-Leyraud}}, \bibinfo {author} {\bibfnamefont {J.}~\bibnamefont
  {Levallois}}, \bibinfo {author} {\bibfnamefont {R.}~\bibnamefont {Daou}},
  \bibinfo {author} {\bibfnamefont {J.-B.}\ \bibnamefont {Bonnemaison}},
  \bibinfo {author} {\bibfnamefont {N.~E.}\ \bibnamefont {Hussey}}, \bibinfo
  {author} {\bibfnamefont {L.}~\bibnamefont {Balicas}}, \bibinfo {author}
  {\bibfnamefont {B.~J.}\ \bibnamefont {Ramshaw}}, \bibinfo {author}
  {\bibfnamefont {R.}~\bibnamefont {Liang}}, \bibinfo {author} {\bibfnamefont
  {D.~A.}\ \bibnamefont {Bonn}}, \bibinfo {author} {\bibfnamefont {W.~N.}\
  \bibnamefont {Hardy}}, \bibinfo {author} {\bibfnamefont {S.}~\bibnamefont
  {Adachi}}, \bibinfo {author} {\bibfnamefont {C.}~\bibnamefont {Proust}}, \
  and\ \bibinfo {author} {\bibfnamefont {L.}~\bibnamefont {Taillefer}},\
  }\href@noop {} {\bibfield  {journal} {\bibinfo  {journal} {Nature}\ }\textbf
  {\bibinfo {volume} {450}},\ \bibinfo {pages} {533} (\bibinfo {year}
  {2007})}\BibitemShut {NoStop}%
\bibitem [{\citenamefont {Sebastian}\ \emph {et~al.}(2008)\citenamefont
  {Sebastian}, \citenamefont {Harrison}, \citenamefont {Palm}, \citenamefont
  {Murphy}, \citenamefont {Mielke}, \citenamefont {Liang}, \citenamefont
  {Bonn}, \citenamefont {Hardy},\ and\ \citenamefont {Lonzarich}}]{seba08}%
  \BibitemOpen
  \bibfield  {author} {\bibinfo {author} {\bibfnamefont {S.~E.}\ \bibnamefont
  {Sebastian}}, \bibinfo {author} {\bibfnamefont {N.}~\bibnamefont {Harrison}},
  \bibinfo {author} {\bibfnamefont {E.}~\bibnamefont {Palm}}, \bibinfo {author}
  {\bibfnamefont {T.~P.}\ \bibnamefont {Murphy}}, \bibinfo {author}
  {\bibfnamefont {C.~H.}\ \bibnamefont {Mielke}}, \bibinfo {author}
  {\bibfnamefont {R.}~\bibnamefont {Liang}}, \bibinfo {author} {\bibfnamefont
  {D.~A.}\ \bibnamefont {Bonn}}, \bibinfo {author} {\bibfnamefont {W.~N.}\
  \bibnamefont {Hardy}}, \ and\ \bibinfo {author} {\bibfnamefont {G.~G.}\
  \bibnamefont {Lonzarich}},\ }\href@noop {} {\bibfield  {journal} {\bibinfo
  {journal} {Nature}\ }\textbf {\bibinfo {volume} {454}},\ \bibinfo {pages}
  {200} (\bibinfo {year} {2008})}\BibitemShut {NoStop}%
\bibitem [{\citenamefont {Vojta}()}]{vojt12}%
  \BibitemOpen
  \bibfield  {author} {\bibinfo {author} {\bibfnamefont {M.}~\bibnamefont
  {Vojta}},\ }\href@noop {} {\enquote {\bibinfo {title} {{Stripes and
  electronic quasiparticles in the pseudogap state of cuprate
  superconductors}},}\ }\Eprint {http://arxiv.org/abs/arXiv:1202.1913}
  {arXiv:1202.1913} \BibitemShut {NoStop}%
\bibitem [{\citenamefont {Kivelson}\ \emph {et~al.}(1998)\citenamefont
  {Kivelson}, \citenamefont {Fradkin},\ and\ \citenamefont {Emery}}]{kive98}%
  \BibitemOpen
  \bibfield  {author} {\bibinfo {author} {\bibfnamefont {S.~A.}\ \bibnamefont
  {Kivelson}}, \bibinfo {author} {\bibfnamefont {E.}~\bibnamefont {Fradkin}}, \
  and\ \bibinfo {author} {\bibfnamefont {V.~J.}\ \bibnamefont {Emery}},\
  }\href@noop {} {\bibfield  {journal} {\bibinfo  {journal} {Nature}\ }\textbf
  {\bibinfo {volume} {393}},\ \bibinfo {pages} {550} (\bibinfo {year}
  {1998})}\BibitemShut {NoStop}%
\bibitem [{\citenamefont {Daou}\ \emph {et~al.}(2010)\citenamefont {Daou},
  \citenamefont {Chang}, \citenamefont {LeBoeuf}, \citenamefont
  {Cyr-Choiniere}, \citenamefont {Laliberte}, \citenamefont {Doiron-Leyraud},
  \citenamefont {Ramshaw}, \citenamefont {Liang}, \citenamefont {Bonn},
  \citenamefont {Hardy},\ and\ \citenamefont {Taillefer}}]{daou10}%
  \BibitemOpen
  \bibfield  {author} {\bibinfo {author} {\bibfnamefont {R.}~\bibnamefont
  {Daou}}, \bibinfo {author} {\bibfnamefont {J.}~\bibnamefont {Chang}},
  \bibinfo {author} {\bibfnamefont {D.}~\bibnamefont {LeBoeuf}}, \bibinfo
  {author} {\bibfnamefont {O.}~\bibnamefont {Cyr-Choiniere}}, \bibinfo {author}
  {\bibfnamefont {F.}~\bibnamefont {Laliberte}}, \bibinfo {author}
  {\bibfnamefont {N.}~\bibnamefont {Doiron-Leyraud}}, \bibinfo {author}
  {\bibfnamefont {B.~J.}\ \bibnamefont {Ramshaw}}, \bibinfo {author}
  {\bibfnamefont {R.}~\bibnamefont {Liang}}, \bibinfo {author} {\bibfnamefont
  {D.~A.}\ \bibnamefont {Bonn}}, \bibinfo {author} {\bibfnamefont {W.~N.}\
  \bibnamefont {Hardy}}, \ and\ \bibinfo {author} {\bibfnamefont
  {L.}~\bibnamefont {Taillefer}},\ }\href
  {http://dx.doi.org/10.1038/nature08716} {\bibfield  {journal} {\bibinfo
  {journal} {Nature}\ }\textbf {\bibinfo {volume} {463}},\ \bibinfo {pages}
  {519} (\bibinfo {year} {2010})}\BibitemShut {NoStop}%
\bibitem [{\citenamefont {Lawler}\ \emph {et~al.}(2010)\citenamefont {Lawler},
  \citenamefont {Fujita}, \citenamefont {Lee}, \citenamefont {Schmidt},
  \citenamefont {Kohsaka}, \citenamefont {Kim}, \citenamefont {Eisaki},
  \citenamefont {Uchida}, \citenamefont {Davis}, \citenamefont {Sethna},\ and\
  \citenamefont {Kim}}]{lawl10}%
  \BibitemOpen
  \bibfield  {author} {\bibinfo {author} {\bibfnamefont {M.~J.}\ \bibnamefont
  {Lawler}}, \bibinfo {author} {\bibfnamefont {K.}~\bibnamefont {Fujita}},
  \bibinfo {author} {\bibfnamefont {J.}~\bibnamefont {Lee}}, \bibinfo {author}
  {\bibfnamefont {A.~R.}\ \bibnamefont {Schmidt}}, \bibinfo {author}
  {\bibfnamefont {Y.}~\bibnamefont {Kohsaka}}, \bibinfo {author} {\bibfnamefont
  {C.~K.}\ \bibnamefont {Kim}}, \bibinfo {author} {\bibfnamefont
  {H.}~\bibnamefont {Eisaki}}, \bibinfo {author} {\bibfnamefont
  {S.}~\bibnamefont {Uchida}}, \bibinfo {author} {\bibfnamefont {J.~C.}\
  \bibnamefont {Davis}}, \bibinfo {author} {\bibfnamefont {J.~P.}\ \bibnamefont
  {Sethna}}, \ and\ \bibinfo {author} {\bibfnamefont {E.-A.}\ \bibnamefont
  {Kim}},\ }\href@noop {} {\bibfield  {journal} {\bibinfo  {journal} {Nature}\
  }\textbf {\bibinfo {volume} {466}},\ \bibinfo {pages} {347} (\bibinfo {year}
  {2010})}\BibitemShut {NoStop}%
\bibitem [{\citenamefont {Mesaros}\ \emph {et~al.}(2011)\citenamefont
  {Mesaros}, \citenamefont {Fujita}, \citenamefont {Eisaki}, \citenamefont
  {Uchida}, \citenamefont {Davis}, \citenamefont {Sachdev}, \citenamefont
  {Zaanen}, \citenamefont {Lawler},\ and\ \citenamefont {Kim}}]{mesa11}%
  \BibitemOpen
  \bibfield  {author} {\bibinfo {author} {\bibfnamefont {A.}~\bibnamefont
  {Mesaros}}, \bibinfo {author} {\bibfnamefont {K.}~\bibnamefont {Fujita}},
  \bibinfo {author} {\bibfnamefont {H.}~\bibnamefont {Eisaki}}, \bibinfo
  {author} {\bibfnamefont {S.}~\bibnamefont {Uchida}}, \bibinfo {author}
  {\bibfnamefont {J.~C.}\ \bibnamefont {Davis}}, \bibinfo {author}
  {\bibfnamefont {S.}~\bibnamefont {Sachdev}}, \bibinfo {author} {\bibfnamefont
  {J.}~\bibnamefont {Zaanen}}, \bibinfo {author} {\bibfnamefont {M.~J.}\
  \bibnamefont {Lawler}}, \ and\ \bibinfo {author} {\bibfnamefont {E.-A.}\
  \bibnamefont {Kim}},\ }\href {\doibase 10.1126/science.1201082} {\bibfield
  {journal} {\bibinfo  {journal} {Science}\ }\textbf {\bibinfo {volume}
  {333}},\ \bibinfo {pages} {426} (\bibinfo {year} {2011})}\BibitemShut
  {NoStop}%
\bibitem [{\citenamefont {Howald}\ \emph {et~al.}(2003)\citenamefont {Howald},
  \citenamefont {Eisaki}, \citenamefont {Kaneko}, \citenamefont {Greven},\ and\
  \citenamefont {Kapitulnik}}]{howa03b}%
  \BibitemOpen
  \bibfield  {author} {\bibinfo {author} {\bibfnamefont {C.}~\bibnamefont
  {Howald}}, \bibinfo {author} {\bibfnamefont {H.}~\bibnamefont {Eisaki}},
  \bibinfo {author} {\bibfnamefont {N.}~\bibnamefont {Kaneko}}, \bibinfo
  {author} {\bibfnamefont {M.}~\bibnamefont {Greven}}, \ and\ \bibinfo {author}
  {\bibfnamefont {A.}~\bibnamefont {Kapitulnik}},\ }\href@noop {} {\bibfield
  {journal} {\bibinfo  {journal} {Phys. Rev. B}\ }\textbf {\bibinfo {volume}
  {67}},\ \bibinfo {pages} {014533} (\bibinfo {year} {2003})}\BibitemShut
  {NoStop}%
\bibitem [{\citenamefont {Kohsaka}\ \emph {et~al.}(2007)\citenamefont
  {Kohsaka}, \citenamefont {Taylor}, \citenamefont {Fujita}, \citenamefont
  {Schmidt}, \citenamefont {Lupien}, \citenamefont {Hanaguri}, \citenamefont
  {Azuma}, \citenamefont {Takano}, \citenamefont {Eisaki}, \citenamefont
  {Takagi}, \citenamefont {Uchida},\ and\ \citenamefont {Davis}}]{kohs07}%
  \BibitemOpen
  \bibfield  {author} {\bibinfo {author} {\bibfnamefont {Y.}~\bibnamefont
  {Kohsaka}}, \bibinfo {author} {\bibfnamefont {C.}~\bibnamefont {Taylor}},
  \bibinfo {author} {\bibfnamefont {K.}~\bibnamefont {Fujita}}, \bibinfo
  {author} {\bibfnamefont {A.}~\bibnamefont {Schmidt}}, \bibinfo {author}
  {\bibfnamefont {C.}~\bibnamefont {Lupien}}, \bibinfo {author} {\bibfnamefont
  {T.}~\bibnamefont {Hanaguri}}, \bibinfo {author} {\bibfnamefont
  {M.}~\bibnamefont {Azuma}}, \bibinfo {author} {\bibfnamefont
  {M.}~\bibnamefont {Takano}}, \bibinfo {author} {\bibfnamefont
  {H.}~\bibnamefont {Eisaki}}, \bibinfo {author} {\bibfnamefont
  {H.}~\bibnamefont {Takagi}}, \bibinfo {author} {\bibfnamefont
  {S.}~\bibnamefont {Uchida}}, \ and\ \bibinfo {author} {\bibfnamefont {J.~C.}\
  \bibnamefont {Davis}},\ }\href@noop {} {\bibfield  {journal} {\bibinfo
  {journal} {Science}\ }\textbf {\bibinfo {volume} {315}},\ \bibinfo {pages}
  {1380} (\bibinfo {year} {2007})}\BibitemShut {NoStop}%
\bibitem [{\citenamefont {Parker}\ \emph {et~al.}(2010)\citenamefont {Parker},
  \citenamefont {Aynajian}, \citenamefont {da~Silva~Neto}, \citenamefont
  {Pushp}, \citenamefont {Ono}, \citenamefont {Wen}, \citenamefont {Xu},
  \citenamefont {Gu},\ and\ \citenamefont {Yazdani}}]{park10}%
  \BibitemOpen
  \bibfield  {author} {\bibinfo {author} {\bibfnamefont {C.~V.}\ \bibnamefont
  {Parker}}, \bibinfo {author} {\bibfnamefont {P.}~\bibnamefont {Aynajian}},
  \bibinfo {author} {\bibfnamefont {E.~H.}\ \bibnamefont {da~Silva~Neto}},
  \bibinfo {author} {\bibfnamefont {A.}~\bibnamefont {Pushp}}, \bibinfo
  {author} {\bibfnamefont {S.}~\bibnamefont {Ono}}, \bibinfo {author}
  {\bibfnamefont {J.}~\bibnamefont {Wen}}, \bibinfo {author} {\bibfnamefont
  {Z.}~\bibnamefont {Xu}}, \bibinfo {author} {\bibfnamefont {G.}~\bibnamefont
  {Gu}}, \ and\ \bibinfo {author} {\bibfnamefont {A.}~\bibnamefont {Yazdani}},\
  }\href@noop {} {\bibfield  {journal} {\bibinfo  {journal} {Nature}\ }\textbf
  {\bibinfo {volume} {468}},\ \bibinfo {pages} {677} (\bibinfo {year}
  {2010})}\BibitemShut {NoStop}%
\bibitem [{\citenamefont {Wise}\ \emph {et~al.}(2008)\citenamefont {Wise},
  \citenamefont {Boyer}, \citenamefont {Chatterjee}, \citenamefont {Kondo},
  \citenamefont {Takeuchi}, \citenamefont {Ikuta}, \citenamefont {Wang},\ and\
  \citenamefont {Hudson}}]{wise08}%
  \BibitemOpen
  \bibfield  {author} {\bibinfo {author} {\bibfnamefont {W.~D.}\ \bibnamefont
  {Wise}}, \bibinfo {author} {\bibfnamefont {M.~C.}\ \bibnamefont {Boyer}},
  \bibinfo {author} {\bibfnamefont {K.}~\bibnamefont {Chatterjee}}, \bibinfo
  {author} {\bibfnamefont {T.}~\bibnamefont {Kondo}}, \bibinfo {author}
  {\bibfnamefont {T.}~\bibnamefont {Takeuchi}}, \bibinfo {author}
  {\bibfnamefont {H.}~\bibnamefont {Ikuta}}, \bibinfo {author} {\bibfnamefont
  {Y.}~\bibnamefont {Wang}}, \ and\ \bibinfo {author} {\bibfnamefont {E.~W.}\
  \bibnamefont {Hudson}},\ }\href@noop {} {\bibfield  {journal} {\bibinfo
  {journal} {Nat. Phys.}\ }\textbf {\bibinfo {volume} {4}},\ \bibinfo {pages}
  {696} (\bibinfo {year} {2008})}\BibitemShut {NoStop}%
\bibitem [{\citenamefont {Fujita}\ \emph {et~al.}(2004)\citenamefont {Fujita},
  \citenamefont {Goka}, \citenamefont {Yamada}, \citenamefont {Tranquada},\
  and\ \citenamefont {Regnault}}]{fuji04}%
  \BibitemOpen
  \bibfield  {author} {\bibinfo {author} {\bibfnamefont {M.}~\bibnamefont
  {Fujita}}, \bibinfo {author} {\bibfnamefont {H.}~\bibnamefont {Goka}},
  \bibinfo {author} {\bibfnamefont {K.}~\bibnamefont {Yamada}}, \bibinfo
  {author} {\bibfnamefont {J.~M.}\ \bibnamefont {Tranquada}}, \ and\ \bibinfo
  {author} {\bibfnamefont {L.~P.}\ \bibnamefont {Regnault}},\ }\href@noop {}
  {\bibfield  {journal} {\bibinfo  {journal} {Phys. Rev. B}\ }\textbf {\bibinfo
  {volume} {70}},\ \bibinfo {pages} {104517} (\bibinfo {year}
  {2004})}\BibitemShut {NoStop}%
\bibitem [{\citenamefont {H\"ucker}\ \emph {et~al.}(2011)\citenamefont
  {H\"ucker}, \citenamefont {v.~Zimmermann}, \citenamefont {Gu}, \citenamefont
  {Xu}, \citenamefont {Wen}, \citenamefont {Xu}, \citenamefont {Kang},
  \citenamefont {Zheludev},\ and\ \citenamefont {Tranquada}}]{huck11}%
  \BibitemOpen
  \bibfield  {author} {\bibinfo {author} {\bibfnamefont {M.}~\bibnamefont
  {H\"ucker}}, \bibinfo {author} {\bibfnamefont {M.}~\bibnamefont
  {v.~Zimmermann}}, \bibinfo {author} {\bibfnamefont {G.~D.}\ \bibnamefont
  {Gu}}, \bibinfo {author} {\bibfnamefont {Z.~J.}\ \bibnamefont {Xu}}, \bibinfo
  {author} {\bibfnamefont {J.~S.}\ \bibnamefont {Wen}}, \bibinfo {author}
  {\bibfnamefont {G.}~\bibnamefont {Xu}}, \bibinfo {author} {\bibfnamefont
  {H.~J.}\ \bibnamefont {Kang}}, \bibinfo {author} {\bibfnamefont
  {A.}~\bibnamefont {Zheludev}}, \ and\ \bibinfo {author} {\bibfnamefont
  {J.~M.}\ \bibnamefont {Tranquada}},\ }\href@noop {} {\bibfield  {journal}
  {\bibinfo  {journal} {Phys. Rev. B}\ }\textbf {\bibinfo {volume} {83}},\
  \bibinfo {pages} {104506} (\bibinfo {year} {2011})}\BibitemShut {NoStop}%
\bibitem [{\citenamefont {Yamada}\ \emph {et~al.}(1998)\citenamefont {Yamada},
  \citenamefont {Lee}, \citenamefont {Kurahashi}, \citenamefont {Wada},
  \citenamefont {Wakimoto}, \citenamefont {Ueki}, \citenamefont {Kimura},
  \citenamefont {Endoh}, \citenamefont {Hosoya}, \citenamefont {Shirane},
  \citenamefont {Birgeneau}, \citenamefont {Greven}, \citenamefont {Kastner},\
  and\ \citenamefont {Kim}}]{yama98a}%
  \BibitemOpen
  \bibfield  {author} {\bibinfo {author} {\bibfnamefont {K.}~\bibnamefont
  {Yamada}}, \bibinfo {author} {\bibfnamefont {C.~H.}\ \bibnamefont {Lee}},
  \bibinfo {author} {\bibfnamefont {K.}~\bibnamefont {Kurahashi}}, \bibinfo
  {author} {\bibfnamefont {J.}~\bibnamefont {Wada}}, \bibinfo {author}
  {\bibfnamefont {S.}~\bibnamefont {Wakimoto}}, \bibinfo {author}
  {\bibfnamefont {S.}~\bibnamefont {Ueki}}, \bibinfo {author} {\bibfnamefont
  {Y.}~\bibnamefont {Kimura}}, \bibinfo {author} {\bibfnamefont
  {Y.}~\bibnamefont {Endoh}}, \bibinfo {author} {\bibfnamefont
  {S.}~\bibnamefont {Hosoya}}, \bibinfo {author} {\bibfnamefont
  {G.}~\bibnamefont {Shirane}}, \bibinfo {author} {\bibfnamefont {R.~J.}\
  \bibnamefont {Birgeneau}}, \bibinfo {author} {\bibfnamefont {M.}~\bibnamefont
  {Greven}}, \bibinfo {author} {\bibfnamefont {M.~A.}\ \bibnamefont {Kastner}},
  \ and\ \bibinfo {author} {\bibfnamefont {Y.~J.}\ \bibnamefont {Kim}},\
  }\href@noop {} {\bibfield  {journal} {\bibinfo  {journal} {Phys. Rev. B}\
  }\textbf {\bibinfo {volume} {57}},\ \bibinfo {pages} {6165} (\bibinfo {year}
  {1998})}\BibitemShut {NoStop}%
\bibitem [{\citenamefont {Birgeneau}\ \emph {et~al.}(2006)\citenamefont
  {Birgeneau}, \citenamefont {Stock}, \citenamefont {Tranquada},\ and\
  \citenamefont {Yamada}}]{birg06}%
  \BibitemOpen
  \bibfield  {author} {\bibinfo {author} {\bibfnamefont {R.~J.}\ \bibnamefont
  {Birgeneau}}, \bibinfo {author} {\bibfnamefont {C.}~\bibnamefont {Stock}},
  \bibinfo {author} {\bibfnamefont {J.~M.}\ \bibnamefont {Tranquada}}, \ and\
  \bibinfo {author} {\bibfnamefont {K.}~\bibnamefont {Yamada}},\ }\href@noop {}
  {\bibfield  {journal} {\bibinfo  {journal} {J. Phys. Soc. Jpn.}\ }\textbf
  {\bibinfo {volume} {75}},\ \bibinfo {pages} {111003} (\bibinfo {year}
  {2006})}\BibitemShut {NoStop}%
\bibitem [{\citenamefont {Shen}\ \emph {et~al.}(2005)\citenamefont {Shen},
  \citenamefont {Ronning}, \citenamefont {Lu}, \citenamefont {Baumberger},
  \citenamefont {Ingle}, \citenamefont {Lee}, \citenamefont {Meevasana},
  \citenamefont {Kohsaka}, \citenamefont {Azuma}, \citenamefont {Takano},
  \citenamefont {Takagi},\ and\ \citenamefont {Shen}}]{shen05}%
  \BibitemOpen
  \bibfield  {author} {\bibinfo {author} {\bibfnamefont {K.~M.}\ \bibnamefont
  {Shen}}, \bibinfo {author} {\bibfnamefont {F.}~\bibnamefont {Ronning}},
  \bibinfo {author} {\bibfnamefont {D.~H.}\ \bibnamefont {Lu}}, \bibinfo
  {author} {\bibfnamefont {F.}~\bibnamefont {Baumberger}}, \bibinfo {author}
  {\bibfnamefont {N.~J.~C.}\ \bibnamefont {Ingle}}, \bibinfo {author}
  {\bibfnamefont {W.~S.}\ \bibnamefont {Lee}}, \bibinfo {author} {\bibfnamefont
  {W.}~\bibnamefont {Meevasana}}, \bibinfo {author} {\bibfnamefont
  {Y.}~\bibnamefont {Kohsaka}}, \bibinfo {author} {\bibfnamefont
  {M.}~\bibnamefont {Azuma}}, \bibinfo {author} {\bibfnamefont
  {M.}~\bibnamefont {Takano}}, \bibinfo {author} {\bibfnamefont
  {H.}~\bibnamefont {Takagi}}, \ and\ \bibinfo {author} {\bibfnamefont {Z.-X.}\
  \bibnamefont {Shen}},\ }\href@noop {} {\bibfield  {journal} {\bibinfo
  {journal} {Science}\ }\textbf {\bibinfo {volume} {307}},\ \bibinfo {pages}
  {901} (\bibinfo {year} {2005})}\BibitemShut {NoStop}%
\bibitem [{\citenamefont {Luo}\ \emph {et~al.}(2007)\citenamefont {Luo},
  \citenamefont {Fang}, \citenamefont {Mu},\ and\ \citenamefont {Wen}}]{luo07}%
  \BibitemOpen
  \bibfield  {author} {\bibinfo {author} {\bibfnamefont {H.}~\bibnamefont
  {Luo}}, \bibinfo {author} {\bibfnamefont {L.}~\bibnamefont {Fang}}, \bibinfo
  {author} {\bibfnamefont {G.}~\bibnamefont {Mu}}, \ and\ \bibinfo {author}
  {\bibfnamefont {H.-H.}\ \bibnamefont {Wen}},\ }\href {\doibase
  10.1016/j.jcrysgro.2007.02.043} {\bibfield  {journal} {\bibinfo  {journal}
  {J. Cryst. Growth}\ }\textbf {\bibinfo {volume} {305}},\ \bibinfo {pages}
  {222} (\bibinfo {year} {2007})}\BibitemShut {NoStop}%
\bibitem [{\citenamefont {Ando}\ \emph {et~al.}(2000)\citenamefont {Ando},
  \citenamefont {Hanaki}, \citenamefont {Ono}, \citenamefont {Murayama},
  \citenamefont {Segawa}, \citenamefont {Miyamoto},\ and\ \citenamefont
  {Komiya}}]{ando00}%
  \BibitemOpen
  \bibfield  {author} {\bibinfo {author} {\bibfnamefont {Y.}~\bibnamefont
  {Ando}}, \bibinfo {author} {\bibfnamefont {Y.}~\bibnamefont {Hanaki}},
  \bibinfo {author} {\bibfnamefont {S.}~\bibnamefont {Ono}}, \bibinfo {author}
  {\bibfnamefont {T.}~\bibnamefont {Murayama}}, \bibinfo {author}
  {\bibfnamefont {K.}~\bibnamefont {Segawa}}, \bibinfo {author} {\bibfnamefont
  {N.}~\bibnamefont {Miyamoto}}, \ and\ \bibinfo {author} {\bibfnamefont
  {S.}~\bibnamefont {Komiya}},\ }\href {\doibase 10.1103/PhysRevB.61.R14956}
  {\bibfield  {journal} {\bibinfo  {journal} {Phys. Rev. B}\ }\textbf {\bibinfo
  {volume} {61}},\ \bibinfo {pages} {R14956} (\bibinfo {year}
  {2000})}\BibitemShut {NoStop}%
\bibitem [{\citenamefont {Pan}\ \emph {et~al.}(2009)\citenamefont {Pan},
  \citenamefont {Richard}, \citenamefont {Xu}, \citenamefont {Neupane},
  \citenamefont {Bishay}, \citenamefont {Fedorov}, \citenamefont {Luo},
  \citenamefont {Fang}, \citenamefont {Wen}, \citenamefont {Wang},\ and\
  \citenamefont {Ding}}]{pan09}%
  \BibitemOpen
  \bibfield  {author} {\bibinfo {author} {\bibfnamefont {Z.-H.}\ \bibnamefont
  {Pan}}, \bibinfo {author} {\bibfnamefont {P.}~\bibnamefont {Richard}},
  \bibinfo {author} {\bibfnamefont {Y.-M.}\ \bibnamefont {Xu}}, \bibinfo
  {author} {\bibfnamefont {M.}~\bibnamefont {Neupane}}, \bibinfo {author}
  {\bibfnamefont {P.}~\bibnamefont {Bishay}}, \bibinfo {author} {\bibfnamefont
  {A.~V.}\ \bibnamefont {Fedorov}}, \bibinfo {author} {\bibfnamefont
  {H.}~\bibnamefont {Luo}}, \bibinfo {author} {\bibfnamefont {L.}~\bibnamefont
  {Fang}}, \bibinfo {author} {\bibfnamefont {H.-H.}\ \bibnamefont {Wen}},
  \bibinfo {author} {\bibfnamefont {Z.}~\bibnamefont {Wang}}, \ and\ \bibinfo
  {author} {\bibfnamefont {H.}~\bibnamefont {Ding}},\ }\href {\doibase
  10.1103/PhysRevB.79.092507} {\bibfield  {journal} {\bibinfo  {journal} {Phys.
  Rev. B}\ }\textbf {\bibinfo {volume} {79}},\ \bibinfo {pages} {092507}
  (\bibinfo {year} {2009})}\BibitemShut {NoStop}%
\bibitem [{\citenamefont {Enoki}\ \emph {et~al.}(2011)\citenamefont {Enoki},
  \citenamefont {Fujita}, \citenamefont {Iikubo}, \citenamefont {Tranquada},\
  and\ \citenamefont {Yamada}}]{enok11}%
  \BibitemOpen
  \bibfield  {author} {\bibinfo {author} {\bibfnamefont {M.}~\bibnamefont
  {Enoki}}, \bibinfo {author} {\bibfnamefont {M.}~\bibnamefont {Fujita}},
  \bibinfo {author} {\bibfnamefont {S.}~\bibnamefont {Iikubo}}, \bibinfo
  {author} {\bibfnamefont {J.~M.}\ \bibnamefont {Tranquada}}, \ and\ \bibinfo
  {author} {\bibfnamefont {K.}~\bibnamefont {Yamada}},\ }\href {\doibase
  10.1143/JPSJS.80SB.SB026} {\bibfield  {journal} {\bibinfo  {journal} {J.
  Phys. Soc. Jpn.}\ }\textbf {\bibinfo {volume} {80}},\ \bibinfo {pages}
  {SB026} (\bibinfo {year} {2011})}\BibitemShut {NoStop}%
\bibitem [{\citenamefont {Hobou}\ \emph {et~al.}(2009)\citenamefont {Hobou},
  \citenamefont {Ishida}, \citenamefont {Fujita}, \citenamefont {Ishikado},
  \citenamefont {Kojima}, \citenamefont {Eisaki},\ and\ \citenamefont
  {Uchida}}]{hobo09}%
  \BibitemOpen
  \bibfield  {author} {\bibinfo {author} {\bibfnamefont {H.}~\bibnamefont
  {Hobou}}, \bibinfo {author} {\bibfnamefont {S.}~\bibnamefont {Ishida}},
  \bibinfo {author} {\bibfnamefont {K.}~\bibnamefont {Fujita}}, \bibinfo
  {author} {\bibfnamefont {M.}~\bibnamefont {Ishikado}}, \bibinfo {author}
  {\bibfnamefont {K.~M.}\ \bibnamefont {Kojima}}, \bibinfo {author}
  {\bibfnamefont {H.}~\bibnamefont {Eisaki}}, \ and\ \bibinfo {author}
  {\bibfnamefont {S.}~\bibnamefont {Uchida}},\ }\href {\doibase
  10.1103/PhysRevB.79.064507} {\bibfield  {journal} {\bibinfo  {journal} {Phys.
  Rev. B}\ }\textbf {\bibinfo {volume} {79}},\ \bibinfo {pages} {064507}
  (\bibinfo {year} {2009})}\BibitemShut {NoStop}%
\bibitem [{\citenamefont {Yamamoto}\ \emph {et~al.}(1998)\citenamefont
  {Yamamoto}, \citenamefont {Katsufuji}, \citenamefont {Tanabe},\ and\
  \citenamefont {Tokura}}]{yama98}%
  \BibitemOpen
  \bibfield  {author} {\bibinfo {author} {\bibfnamefont {K.}~\bibnamefont
  {Yamamoto}}, \bibinfo {author} {\bibfnamefont {T.}~\bibnamefont {Katsufuji}},
  \bibinfo {author} {\bibfnamefont {T.}~\bibnamefont {Tanabe}}, \ and\ \bibinfo
  {author} {\bibfnamefont {Y.}~\bibnamefont {Tokura}},\ }\href@noop {}
  {\bibfield  {journal} {\bibinfo  {journal} {Phys. Rev. Lett.}\ }\textbf
  {\bibinfo {volume} {80}},\ \bibinfo {pages} {1493} (\bibinfo {year}
  {1998})}\BibitemShut {NoStop}%
\bibitem [{\citenamefont {Dai}\ \emph {et~al.}(2001)\citenamefont {Dai},
  \citenamefont {Mook}, \citenamefont {Hunt},\ and\ \citenamefont
  {Do\u{g}an}}]{dai01}%
  \BibitemOpen
  \bibfield  {author} {\bibinfo {author} {\bibfnamefont {P.}~\bibnamefont
  {Dai}}, \bibinfo {author} {\bibfnamefont {H.~A.}\ \bibnamefont {Mook}},
  \bibinfo {author} {\bibfnamefont {R.~D.}\ \bibnamefont {Hunt}}, \ and\
  \bibinfo {author} {\bibfnamefont {F.}~\bibnamefont {Do\u{g}an}},\ }\href@noop
  {} {\bibfield  {journal} {\bibinfo  {journal} {Phys. Rev. B}\ }\textbf
  {\bibinfo {volume} {63}},\ \bibinfo {pages} {054525} (\bibinfo {year}
  {2001})}\BibitemShut {NoStop}%
\bibitem [{\citenamefont {Haug}\ \emph {et~al.}(2010)\citenamefont {Haug},
  \citenamefont {Hinkov}, \citenamefont {Sidis}, \citenamefont {Bourges},
  \citenamefont {Christensen}, \citenamefont {Ivanov}, \citenamefont {Keller},
  \citenamefont {Lin},\ and\ \citenamefont {Keimer}}]{haug10}%
  \BibitemOpen
  \bibfield  {author} {\bibinfo {author} {\bibfnamefont {D.}~\bibnamefont
  {Haug}}, \bibinfo {author} {\bibfnamefont {V.}~\bibnamefont {Hinkov}},
  \bibinfo {author} {\bibfnamefont {Y.}~\bibnamefont {Sidis}}, \bibinfo
  {author} {\bibfnamefont {P.}~\bibnamefont {Bourges}}, \bibinfo {author}
  {\bibfnamefont {N.~B.}\ \bibnamefont {Christensen}}, \bibinfo {author}
  {\bibfnamefont {A.}~\bibnamefont {Ivanov}}, \bibinfo {author} {\bibfnamefont
  {T.}~\bibnamefont {Keller}}, \bibinfo {author} {\bibfnamefont {C.~T.}\
  \bibnamefont {Lin}}, \ and\ \bibinfo {author} {\bibfnamefont
  {B.}~\bibnamefont {Keimer}},\ }\href@noop {} {\bibfield  {journal} {\bibinfo
  {journal} {New J. Phys.}\ }\textbf {\bibinfo {volume} {12}},\ \bibinfo
  {pages} {105006} (\bibinfo {year} {2010})}\BibitemShut {NoStop}%
\bibitem [{\citenamefont {Enoki}\ \emph {et~al.}(2010)\citenamefont {Enoki},
  \citenamefont {Fujita}, \citenamefont {Iikubo},\ and\ \citenamefont
  {Yamada}}]{enok10}%
  \BibitemOpen
  \bibfield  {author} {\bibinfo {author} {\bibfnamefont {M.}~\bibnamefont
  {Enoki}}, \bibinfo {author} {\bibfnamefont {M.}~\bibnamefont {Fujita}},
  \bibinfo {author} {\bibfnamefont {S.}~\bibnamefont {Iikubo}}, \ and\ \bibinfo
  {author} {\bibfnamefont {K.}~\bibnamefont {Yamada}},\ }\href {\doibase
  10.1016/j.physc.2009.11.010} {\bibfield  {journal} {\bibinfo  {journal}
  {Physica C}\ }\textbf {\bibinfo {volume} {470}},\ \bibinfo {pages} {S37}
  (\bibinfo {year} {2010})}\BibitemShut {NoStop}%
\bibitem [{\citenamefont {Wakimoto}\ \emph {et~al.}(2000)\citenamefont
  {Wakimoto}, \citenamefont {Birgeneau}, \citenamefont {Kastner}, \citenamefont
  {Lee}, \citenamefont {Erwin}, \citenamefont {Gehring}, \citenamefont {Lee},
  \citenamefont {Fujita}, \citenamefont {Yamada}, \citenamefont {Endoh},
  \citenamefont {Hirota},\ and\ \citenamefont {Shirane}}]{waki00}%
  \BibitemOpen
  \bibfield  {author} {\bibinfo {author} {\bibfnamefont {S.}~\bibnamefont
  {Wakimoto}}, \bibinfo {author} {\bibfnamefont {R.~J.}\ \bibnamefont
  {Birgeneau}}, \bibinfo {author} {\bibfnamefont {M.~A.}\ \bibnamefont
  {Kastner}}, \bibinfo {author} {\bibfnamefont {Y.~S.}\ \bibnamefont {Lee}},
  \bibinfo {author} {\bibfnamefont {R.}~\bibnamefont {Erwin}}, \bibinfo
  {author} {\bibfnamefont {P.~M.}\ \bibnamefont {Gehring}}, \bibinfo {author}
  {\bibfnamefont {S.~H.}\ \bibnamefont {Lee}}, \bibinfo {author} {\bibfnamefont
  {M.}~\bibnamefont {Fujita}}, \bibinfo {author} {\bibfnamefont
  {K.}~\bibnamefont {Yamada}}, \bibinfo {author} {\bibfnamefont
  {Y.}~\bibnamefont {Endoh}}, \bibinfo {author} {\bibfnamefont
  {K.}~\bibnamefont {Hirota}}, \ and\ \bibinfo {author} {\bibfnamefont
  {G.}~\bibnamefont {Shirane}},\ }\href@noop {} {\bibfield  {journal} {\bibinfo
   {journal} {Phys. Rev. B}\ }\textbf {\bibinfo {volume} {61}},\ \bibinfo
  {pages} {3699} (\bibinfo {year} {2000})}\BibitemShut {NoStop}%
\bibitem [{\citenamefont {Fujita}\ \emph {et~al.}(2002)\citenamefont {Fujita},
  \citenamefont {Yamada}, \citenamefont {Hiraka}, \citenamefont {Gehring},
  \citenamefont {Lee}, \citenamefont {Wakimoto},\ and\ \citenamefont
  {Shirane}}]{fuji02c}%
  \BibitemOpen
  \bibfield  {author} {\bibinfo {author} {\bibfnamefont {M.}~\bibnamefont
  {Fujita}}, \bibinfo {author} {\bibfnamefont {K.}~\bibnamefont {Yamada}},
  \bibinfo {author} {\bibfnamefont {H.}~\bibnamefont {Hiraka}}, \bibinfo
  {author} {\bibfnamefont {P.~M.}\ \bibnamefont {Gehring}}, \bibinfo {author}
  {\bibfnamefont {S.~H.}\ \bibnamefont {Lee}}, \bibinfo {author} {\bibfnamefont
  {S.}~\bibnamefont {Wakimoto}}, \ and\ \bibinfo {author} {\bibfnamefont
  {G.}~\bibnamefont {Shirane}},\ }\href@noop {} {\bibfield  {journal} {\bibinfo
   {journal} {Phys. Rev. B}\ }\textbf {\bibinfo {volume} {65}},\ \bibinfo
  {pages} {064505} (\bibinfo {year} {2002})}\BibitemShut {NoStop}%
\bibitem [{\citenamefont {Luo}\ \emph {et~al.}(2008)\citenamefont {Luo},
  \citenamefont {Cheng}, \citenamefont {Fang},\ and\ \citenamefont
  {Wen}}]{luo08}%
  \BibitemOpen
  \bibfield  {author} {\bibinfo {author} {\bibfnamefont {H.}~\bibnamefont
  {Luo}}, \bibinfo {author} {\bibfnamefont {P.}~\bibnamefont {Cheng}}, \bibinfo
  {author} {\bibfnamefont {L.}~\bibnamefont {Fang}}, \ and\ \bibinfo {author}
  {\bibfnamefont {H.-H.}\ \bibnamefont {Wen}},\ }\href@noop {} {\bibfield
  {journal} {\bibinfo  {journal} {Supercond. Sci. Technol.}\ }\textbf {\bibinfo
  {volume} {21}},\ \bibinfo {pages} {125024} (\bibinfo {year}
  {2008})}\BibitemShut {NoStop}%
\bibitem [{\citenamefont {Hashimoto}\ \emph {et~al.}(2008)\citenamefont
  {Hashimoto}, \citenamefont {Yoshida}, \citenamefont {Yagi}, \citenamefont
  {Takizawa}, \citenamefont {Fujimori}, \citenamefont {Kubota}, \citenamefont
  {Ono}, \citenamefont {Tanaka}, \citenamefont {Lu}, \citenamefont {Shen},
  \citenamefont {Ono},\ and\ \citenamefont {Ando}}]{hash08}%
  \BibitemOpen
  \bibfield  {author} {\bibinfo {author} {\bibfnamefont {M.}~\bibnamefont
  {Hashimoto}}, \bibinfo {author} {\bibfnamefont {T.}~\bibnamefont {Yoshida}},
  \bibinfo {author} {\bibfnamefont {H.}~\bibnamefont {Yagi}}, \bibinfo {author}
  {\bibfnamefont {M.}~\bibnamefont {Takizawa}}, \bibinfo {author}
  {\bibfnamefont {A.}~\bibnamefont {Fujimori}}, \bibinfo {author}
  {\bibfnamefont {M.}~\bibnamefont {Kubota}}, \bibinfo {author} {\bibfnamefont
  {K.}~\bibnamefont {Ono}}, \bibinfo {author} {\bibfnamefont {K.}~\bibnamefont
  {Tanaka}}, \bibinfo {author} {\bibfnamefont {D.~H.}\ \bibnamefont {Lu}},
  \bibinfo {author} {\bibfnamefont {Z.-X.}\ \bibnamefont {Shen}}, \bibinfo
  {author} {\bibfnamefont {S.}~\bibnamefont {Ono}}, \ and\ \bibinfo {author}
  {\bibfnamefont {Y.}~\bibnamefont {Ando}},\ }\href {\doibase
  10.1103/PhysRevB.77.094516} {\bibfield  {journal} {\bibinfo  {journal} {Phys.
  Rev. B}\ }\textbf {\bibinfo {volume} {77}},\ \bibinfo {pages} {094516}
  (\bibinfo {year} {2008})}\BibitemShut {NoStop}%
\bibitem [{\citenamefont {Sushkov}(2009)}]{sush09}%
  \BibitemOpen
  \bibfield  {author} {\bibinfo {author} {\bibfnamefont {O.~P.}\ \bibnamefont
  {Sushkov}},\ }\href@noop {} {\bibfield  {journal} {\bibinfo  {journal} {Phys.
  Rev. B}\ }\textbf {\bibinfo {volume} {79}},\ \bibinfo {eid} {174519}
  (\bibinfo {year} {2009})}\BibitemShut {NoStop}%
\bibitem [{\citenamefont {Seibold}\ \emph {et~al.}(2011)\citenamefont
  {Seibold}, \citenamefont {Markiewicz},\ and\ \citenamefont
  {Lorenzana}}]{seib11}%
  \BibitemOpen
  \bibfield  {author} {\bibinfo {author} {\bibfnamefont {G.}~\bibnamefont
  {Seibold}}, \bibinfo {author} {\bibfnamefont {R.~S.}\ \bibnamefont
  {Markiewicz}}, \ and\ \bibinfo {author} {\bibfnamefont {J.}~\bibnamefont
  {Lorenzana}},\ }\href {\doibase 10.1103/PhysRevB.83.205108} {\bibfield
  {journal} {\bibinfo  {journal} {Phys. Rev. B}\ }\textbf {\bibinfo {volume}
  {83}},\ \bibinfo {pages} {205108} (\bibinfo {year} {2011})}\BibitemShut
  {NoStop}%
\bibitem [{\citenamefont {Lee}\ \emph {et~al.}(2000)\citenamefont {Lee},
  \citenamefont {Yamada}, \citenamefont {Endoh}, \citenamefont {Shirane},
  \citenamefont {Birgeneau}, \citenamefont {Kastner}, \citenamefont {Greven},\
  and\ \citenamefont {Kim}}]{lee00}%
  \BibitemOpen
  \bibfield  {author} {\bibinfo {author} {\bibfnamefont {C.-H.}\ \bibnamefont
  {Lee}}, \bibinfo {author} {\bibfnamefont {K.}~\bibnamefont {Yamada}},
  \bibinfo {author} {\bibfnamefont {Y.}~\bibnamefont {Endoh}}, \bibinfo
  {author} {\bibfnamefont {G.}~\bibnamefont {Shirane}}, \bibinfo {author}
  {\bibfnamefont {R.~J.}\ \bibnamefont {Birgeneau}}, \bibinfo {author}
  {\bibfnamefont {M.~A.}\ \bibnamefont {Kastner}}, \bibinfo {author}
  {\bibfnamefont {M.}~\bibnamefont {Greven}}, \ and\ \bibinfo {author}
  {\bibfnamefont {Y.-J.}\ \bibnamefont {Kim}},\ }\href@noop {} {\bibfield
  {journal} {\bibinfo  {journal} {J. Phys. Soc. Jpn.}\ }\textbf {\bibinfo
  {volume} {69}},\ \bibinfo {pages} {1170} (\bibinfo {year}
  {2000})}\BibitemShut {NoStop}%
\bibitem [{\citenamefont {Meng}\ \emph {et~al.}(2011)\citenamefont {Meng},
  \citenamefont {Brunner}, \citenamefont {Kim}, \citenamefont {Lee},
  \citenamefont {Lee}, \citenamefont {Wen}, \citenamefont {Xu}, \citenamefont
  {Gu},\ and\ \citenamefont {Gweon}}]{gweo11}%
  \BibitemOpen
  \bibfield  {author} {\bibinfo {author} {\bibfnamefont {J.-Q.}\ \bibnamefont
  {Meng}}, \bibinfo {author} {\bibfnamefont {M.}~\bibnamefont {Brunner}},
  \bibinfo {author} {\bibfnamefont {K.-H.}\ \bibnamefont {Kim}}, \bibinfo
  {author} {\bibfnamefont {H.-G.}\ \bibnamefont {Lee}}, \bibinfo {author}
  {\bibfnamefont {S.-I.}\ \bibnamefont {Lee}}, \bibinfo {author} {\bibfnamefont
  {J.~S.}\ \bibnamefont {Wen}}, \bibinfo {author} {\bibfnamefont {Z.~J.}\
  \bibnamefont {Xu}}, \bibinfo {author} {\bibfnamefont {G.~D.}\ \bibnamefont
  {Gu}}, \ and\ \bibinfo {author} {\bibfnamefont {G.-H.}\ \bibnamefont
  {Gweon}},\ }\href@noop {} {\bibfield  {journal} {\bibinfo  {journal} {Phys.
  Rev. B}\ }\textbf {\bibinfo {volume} {84}},\ \bibinfo {pages} {060513}
  (\bibinfo {year} {2011})}\BibitemShut {NoStop}%
\bibitem [{\citenamefont {Stock}\ \emph {et~al.}(2010)\citenamefont {Stock},
  \citenamefont {Cowley}, \citenamefont {Buyers}, \citenamefont {Frost},
  \citenamefont {Taylor}, \citenamefont {Peets}, \citenamefont {Liang},
  \citenamefont {Bonn},\ and\ \citenamefont {Hardy}}]{stoc10}%
  \BibitemOpen
  \bibfield  {author} {\bibinfo {author} {\bibfnamefont {C.}~\bibnamefont
  {Stock}}, \bibinfo {author} {\bibfnamefont {R.~A.}\ \bibnamefont {Cowley}},
  \bibinfo {author} {\bibfnamefont {W.~J.~L.}\ \bibnamefont {Buyers}}, \bibinfo
  {author} {\bibfnamefont {C.~D.}\ \bibnamefont {Frost}}, \bibinfo {author}
  {\bibfnamefont {J.~W.}\ \bibnamefont {Taylor}}, \bibinfo {author}
  {\bibfnamefont {D.}~\bibnamefont {Peets}}, \bibinfo {author} {\bibfnamefont
  {R.}~\bibnamefont {Liang}}, \bibinfo {author} {\bibfnamefont
  {D.}~\bibnamefont {Bonn}}, \ and\ \bibinfo {author} {\bibfnamefont {W.~N.}\
  \bibnamefont {Hardy}},\ }\href@noop {} {\bibfield  {journal} {\bibinfo
  {journal} {Phys. Rev. B}\ }\textbf {\bibinfo {volume} {82}},\ \bibinfo
  {pages} {174505} (\bibinfo {year} {2010})}\BibitemShut {NoStop}%
\end{thebibliography}%

\end{document}